\documentclass{article}
\usepackage{spconf,amsmath,graphicx}

\usepackage{booktabs}
\usepackage{graphicx}
\usepackage{xcolor}
\usepackage{amsmath, amssymb, graphicx}

\definecolor{green1}{HTML}{00684D}
\definecolor{purple1}{HTML}{002D2E}
\definecolor{pink}{HTML}{fc0377}

\usepackage{algorithm}
\usepackage{algpseudocode}
\usepackage{url}
\usepackage{ragged2e}
\usepackage{balance}

\title{An Investigation of Incorporating Mamba for Speech Enhancement}
\name{\begin{tabular}{@{}c@{}}Rong Chao$^\ddagger$$^{\S}$, Wen-Huang Cheng$^\S$, Moreno La Quatra$^\dagger$ \\ Sabato Marco Siniscalchi$^+$, \textit{Chao-Han Huck Yang$^*$, Szu-Wei Fu$^*$, Yu Tsao$^\ddagger$}\end{tabular}}
\address{$^\ddagger$Academia Sinica, $^\S$National Taiwan University, $^\dagger$Kore University of Enna,\\ $^+$University of Palermo, $^*$NVIDIA}

\begin{document}
%\ninept
%
\maketitle
\begin{abstract}
This work aims to investigate the use of a recently proposed, attention-free, scalable state-space model (SSM), Mamba, for the speech enhancement (SE) task. In particular, we employ Mamba to deploy different regression-based SE models (SEMamba) with different configurations, namely basic, advanced, causal, and non-causal. Furthermore, loss functions either based on signal-level distances or metric-oriented are considered. Experimental evidence shows that SEMamba attains a competitive PESQ of 3.55 on the VoiceBank-DEMAND dataset with the advanced, non-causal configuration. A new state-of-the-art PESQ of 3.69 is also reported when SEMamba is combined with Perceptual Contrast Stretching (PCS). Compared against Transformed-based equivalent SE solutions, a noticeable FLOPs reduction up to $\sim$12\% is observed with the advanced non-causal configurations. Finally, SEMamba can be used as a pre-processing step before automatic speech recognition (ASR), showing competitive performance against recent SE solutions. 
\end{abstract}

\begin{figure*}[htb]
    \centering
    % \centerline{\includegraphics[width=14cm]{figures/simpleMamba.pdf}}
    % \centerline{\includegraphics[width=18.1cm]{figures/simpleMamba.pdf}}
    \centerline{\includegraphics[width=18.1cm]{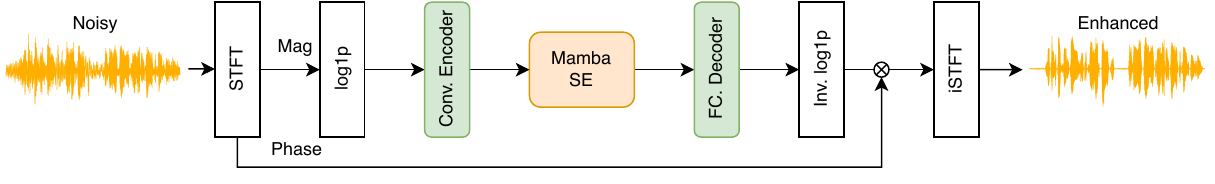}}
    \caption{Architecture of our basic Mamba-based Speech Enhancement (SE) model, SEMamba-basic.}
    \label{fig:causal_se}
    % \vspace{-5mm}
\end{figure*}

\section{Introduction}
\label{sec:intro}

A speech enhancement (SE) process involves retrieving clean speech components from a distorted signal to generate an improved version with enhanced acoustic properties~\cite{SE, wang2018supervised}. 
In various speech applications, such as assistive hearing technologies~\cite{HA, CI}, speaker recognition~\cite{SR, michelsanti2017conditional}, and automatic speech recognition (ASR)~\cite{ochiai2024rethinking, ASR}, SE serves as a critical front-end processor. Recently, SE has been framed as a regression task utilizing neural-network-based mapping functions to convert input noisy speech signals into cleaner outputs. Specifically, deep neural models are employed to implement the mapping function that enhances noisy speech. Several deep architectures, including deep denoising autoencoders~\cite{DAELu}, deep fully connected neural networks~\cite{SEDNN, liu2014experiments}, convolutional neural networks~\cite{FCN}, long short-term memory~\cite{LSTM}, convolutional recurrent neural networks~\cite{li2020speech,
tan2018convolutional}, and more recently, Transformer and its variant Conformer, have been used to form the mapping function, resulting in notably promising SE performance~\cite{Trans, CMGAN, mp_senet, lu2023explicit}.

In addition to exploring advanced model architectures, several studies have aimed to develop effective objective functions to boost SE capability. Typically, a signal-level distance measure serves as the fundamental objective function during training, e.g.,  L1/L2 norm~\cite{qi2020mean}, SI-SDR~\cite{SDR}, or multiple-resolution loss~\cite{demucs_stft}. Objective functions specifically crafted and optimized for the downstream SE tasks have also been proposed. For instance, objective scores, such as perceptual evaluation of speech quality (PESQ) ~\cite{Rix2001}, or short-time objective intelligibility (STOI) ~\cite{Taal2011}, are learned using neural network models in~\cite{fu2019metricgan, fu2021metricgan+} and then employed to optimize the SE model. Similarly, in~\cite{nayem2023attention}, subjective assessment results are learned through a quality estimator and utilized for training the SE model. Furthermore, a few studies have shown improved SE results by preprocessing the target speech prior to model training. For example, perceptual contrast stretching (PCS)~\cite{chao2022perceptual} enhances the contrast of target features based on their perceptual importance, effectively boosting SE performance without a notable increase in computational cost during runtime.

Recently, a highly promising neural architecture, referred to as Mamba~\cite{gu2023mamba}, has emerged, harnessing the state-space model with a novel selection mechanism. Mamba exhibits comparable or superior performance to state-of-the-art (SOTA) Transformer-based models across diverse tasks \cite{jiang2024dual, li2024spmamba} due to its intrinsic capability in modeling extremely long-range dependencies. 
Mamba stands out for its efficient use of computational resources, scaling linearly in sequence length compared to the quadratic complexity of Transformers, making it much more resource-efficient for long sequences. 

In this work, we are interested in comparing and contrasting Transformer- and Mamba-based SE solutions, and we thereby introduce a new Mamba-based SE solution, named SEMamba. We evaluate different SEMamba configuration: first, we deploy a basic SE configuration, termed SEMamba-basic. Next, we implement SEMamba with an advanced configuration, termed SEMamba-advanced. Both causal, and non-causal solutions are tested along with conventional distance-based and metric-oriented losses.
We assess our SEMamba on the VoiceBank-DEMAND dataset~\cite{valentini2016investigating}, and contrast it with transformer-based counterparts. Experimental results demonstrate that (i) SEMamba attains a competitive PESQ of 3.55 on the VoiceBank-DEMAND dataset with the advanced, non-causal configuration, and (ii) a new state-of-the-art PESQ of 3.69 is delivered when  combined with PCS. Compared with Transformed-based basic architecture,  a reduction in the number of parameters up to $\sim$60\% and $\sim$28\% is also observed along with a floating-point operations per second (FLOPs) reduction up to $\sim$66\% and $\sim$53\% with causal and non-causal configurations, respectively. In the advanced configuration,  a notable FLOPs reduction up to $\sim$12\% is observed when replacing Transformers with Mamba, while keeping the similar amount of parameters. Finally, SEMamba can be used as a pre-processing step before ASR, showing competitive performance against recent SE solutions.

\section{Mamba: Linear-Time Sequence Modeling with Selective State Spaces}
\label{sec:model}
The structured state space model (SSM)~\cite{gu2021efficiently} has demonstrated the ability to manage long-dependent sequences with low computation and memory requirements. 
It can serve as a replacement for either a CNN for efficient parallel training or an RNN for rapid autoregressive generation~\cite{gu2022parameterization}. 
More recently, Mamba~\cite{gu2023mamba} has introduced notable advancements in discrete data modeling of SSM by introducing two key improvements. First, Mamba incorporates an input-dependent selection mechanism, enabling efficient information filtering by parameterizing the SSM module based on the input information. Second, Mamba introduces a hardware-aware algorithm that scales linearly with input sequence length, facilitating faster computation of the model recurrently with a scan. The Mamba architecture, which integrates SSM blocks with linear layers, is notably simpler than its Transformer-based counterparts and has demonstrated SOTA performance across various long-sequence patterns, including language and genomics. Therefore, noticeable computational efficiency during both training and inference phases has been attained.

Structured SSMs, as described in the Mamba method~\cite{gu2023mamba}, operate by mapping an input \textbf{\textit{x}} to an output \textbf{\textit{y}} through a higher  dimensional latent state \textbf{\textit{h}}, as follows: $ h_n = \bar{\textbf{\textit{A}}}h_{n-1} + \bar{\textbf{\textit{B}}}x_n$, and $y_n = \textbf{\textit{C}}h_n$. \(\bar{\textbf{\textit{A}}}\) and \(\bar{\textbf{\textit{B}}}\) represent discretized state matrices. The discretization process converts continuous parameters \((\bf{\Delta}, \textbf{\textit{A}}, \textbf{\textit{B}})\) into discrete counterparts \((\bar{\textbf{\textit{A}}}, \bar{\textbf{\textit{B}}})\).
Mamba integrates components from the H3 architecture~\cite{fu2022hungry} and a gated multilayer perceptron block into a stacked structure, thereby expanding the model's internal representation dimension and concentrating most parameters in linear projections. %thereby expanding the model dimension by a factor and focusing most parameters in linear projections. 
This design results in fewer parameters for the inner SSM and adopts the SiLU/Swish activation function along with standard normalization and residual connections. For further details, the interested reader is referred to~\cite{gu2023mamba}.

%\begin{align}
%h_n &= \bar{\textbf{\textit{A}}}h_{n-1} + \bar{\textbf{\textit{B}}}x_n, \tag{1} \\
%y_n &= \textbf{\textit{C}}h_n \tag{2}
%\end{align}
%where \(\bar{\textbf{\textit{A}}}\) and \(\bar{\textbf{\textit{B}}}\) represent discretized state matrices. The discretization process converts continuous parameters \((\bf{\Delta}, \textbf{\textit{A}}, \textbf{\textit{B}})\) into discrete counterparts \((\bar{\textbf{\textit{A}}}, \bar{\textbf{\textit{B}}})\).%, allowing the model to process discrete-time audio signals~\cite{gu2023mamba}.

\section{Mamba in Speech Enhancement}

\begin{figure*}[htb]
    \centering
    
    % \centerline{\includegraphics[width=15cm]{figures/SEMamba.pdf}}
    % \centerline{\includegraphics[width=18.1cm]{figures/SEMamba.pdf}}
    \centerline{\includegraphics[width=18.1cm]{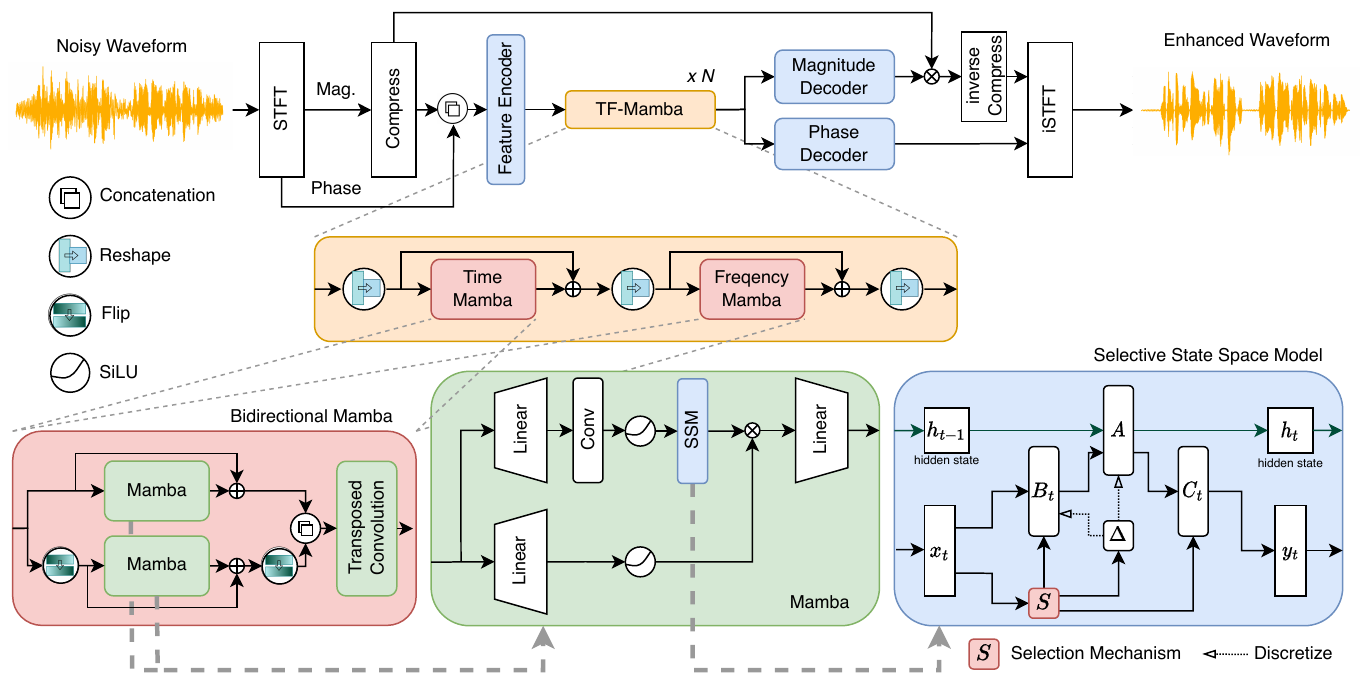}}
    \caption{Architecture of the proposed SEMamba-advanced with Time-Frequency (TF) and Selective-SSM mechanism.}
    \label{fig:architecture}
    % \vspace{-4mm}
\end{figure*}

Two SEMamba architectures are deployed: The first architecture integrates Mamba in a basic SE architecture (see Fig. \ref{fig:causal_se}). 
The second architecture leverages MP-SENet but uses the Mamba block in place of attention-based ones (see in Fig. \ref{fig:architecture}). %More details are given in the following sections.

\subsection{SEMamba-basic}
%We first explore the construction of SEMamba using a basic causal model architecture.
SEMamba-basic is  causal, i.e., the output at time $t$, $y_t$, is influenced only by preceding inputs $x_n$ with $n \in \{0, \dots, t\}$. %- which implies a causal SE system. 
As depicted in Fig. \ref{fig:causal_se}, the input noisy waveform is first transformed into its spectral representation using the Short-Time Fourier Transform (STFT). Next, the magnitude (Mag) component of the STFT is compressed using the $log1p$ function $(log1p(z) = log(1+z))$~\cite{fu2020boosting} to regulate the dynamic range of the magnitude. The compressed spectral magnitude is then fed into the SE module, which comprises (i) a convolutional encoder having 4 convolutional layers, (ii) two uni-directional Mamba blocks, and (ii) a fully connected decoder layer. The enhanced  magnitude is decompressed by an inverse of $log1p$ before being combined with the noisy phase in the iSTFT block to reconstruct the enhanced waveform.
We also implement the basic SE architecture with  Transformers, following the architecture used
in~\cite{fu2020boosting}. On the other hand, SEMamba-basic employs the Mamba blocks to replace the Transformer blocks. Both basic models are trained using the mean absolute error of the magnitude as the loss function. 
%Both models are trained using mean absolute error on the magnitude as the loss function.%, allowing us to evaluate the difference between the attention and the selective state mechanisms.

% \vspace{-2mm}
\subsection{SEMamba-advanced}
In the architecture shown in Fig. \ref{fig:architecture}, the causality constraint is removed, and a more sophisticated SE structure is employed. Specifically, MP-SENet~\cite{mp_senet}, which leverages both magnitude and phase spectra, serves as the architectural backbone. A spectral representation is obtained from the input noisy waveform using the STFT. The magnitude component is first compressed and then stacked together with the phase components. These components are processed by a feature encoder, featuring a dilated DenseNet core flanked by two convolutional layers. The output of the feature encoder undergoes multiple transformations via the Time-Frequency Mamba block, which is executed $N$ times ($N$=4 in this study) to enhance its spectral properties. The output of the Time-Frequency Mamba block is then routed to two separate decoders for magnitude and phase processing, respectively. 
Each decoder comprises a dilated DenseNet followed by a deconvolutional block, and it ends with an output layer that employs a 2D-convolution. 
% The output layers reconstruct the magnitude mask and the real and imaginary parts of the waveform, respectively. 
 
 The loss function used to train SEMamba-advanced leverages a linear combination of losses, including PESQ-based GAN discriminator, time, magnitude, complex, and phase losses, as in MP-SENet.
% \vspace{-2mm}
\subsection{SEMamba-advanced \& additional designs}

\subsubsection{From uni- to bi-directional Mamba}
We explore the possible benefit of modifying the Mamba structure  from a uni-directional to a bi-directional configuration. This modification involves adapting the module to process input sequences both in their original and reversed forms. The inputs are processed in parallel using Mamba modules, after which the outputs are concatenated. The output is then fed to a $Conv1D$ layer:

\begin{equation}
\textbf{y} = Conv1D( M_{uni}(\textbf{x}) \oplus flip(M_{uni}(flip(\textbf{x})) )), 
\end{equation}
%\[
%\textbf{y} = Conv1D( M_{uni}(\textbf{x}) \oplus M_{uni}(flip(\textbf{x})) ),
%\]
\noindent where $\textbf{x}$, $\textbf{y}$, $M_{uni}()$, $flip()$, $Conv1D()$, and $\oplus$, respectively, denote input, output, uni-directional Mamba, flipping operation, 1-D convolution, and concatenation.

\subsubsection{Consistency loss (CL)}
% \roychao{Consistancy issue: FAST SIGNAL RECONSTRUCTION FROM MAGNITUDE STFT SPECTROGRAM
% BASED ON SPECTROGRAM CONSISTENCY}

Stability during the training process can be improved by leveraging the consistency loss proposed in \cite{zadorozhnyy2022scp}. The consistency loss aims to minimize the gap between the complex spectrum obtained directly from the model output (comprising amplitude and phase) and the complex spectrum derived after applying inverse Short-Time Fourier Transform (iSTFT) and then re-applying STFT to the resulting waveform. This gap arises because the complex spectrum predicted directly by the model may fall outside the STFT domain. Once processed through iSTFT and then subjected to STFT, the spectrum re-enters the STFT domain, thereby closing this gap. This mechanism ensures that our enhancements in the complex domain translate effectively in the time-frequency domain.

\subsubsection{Perceptual contrast stretching (PCS)}
\label{ssec:pcs} 
PCS is a spectral processing technique that aims to improve the perceptual quality of speech signals. It leverages empirical observations of varying sensitivity levels in the human auditory system. 
PCS exploits this phenomenon by stretching the magnitude spectrum of the signal based on each frequency band's perceived importance.
In this paper, we introduce PCS as an auxiliary step following the enhancement process, aiming to refine the perceptual quality of the speech signal~\cite{chao2022perceptual}.

\begin{table}
\caption{Comparison of basic SE architectures leveraging either Mamba or Transformer blocks.}
\label{table_uni}
% \vspace{0.5em}
\vspace{0.8em}
\centering
\begin{tabular*}{\linewidth}{l||llllll}
% \begin{tabular*}{8cm}{@{\extracolsep{\fill}} l || l l l l l l}
\specialrule{1pt}{1pt}{1pt}
                  & Causal & PESQ  & STOI &  FLOPs & Param. \\ 
\hline
noisy-speech             & -- & 1.97 & 0.92 &  -- &   -- \\
\hline

% Transformer \cite{fu2020boosting}           & Yes & 2.76 & 0.94  & 2.26G & 9.05M  \\
Transf. \cite{fu2020boosting}           & Yes & 2.76 & 0.94  & 2.26G & 9.05M  \\
Mamba           & Yes & 2.76 & 0.94  & \textbf{0.76}G & \textbf{3.60M}   \\
% \hline
% Transformer \cite{fu2020boosting}           & No & 2.84 & 0.94  & 2.26G & 9.05M  \\
Transf. \cite{fu2020boosting}           & No & 2.84 & 0.94  & 2.26G & 9.05M  \\
Mamba           & No & \textbf{2.85} & 0.94  & \textbf{1.06G} & \textbf{6.49M}   \\

\specialrule{1pt}{1pt}{1pt}
\end{tabular*}
\end{table}

\begin{table}
\caption{Comparison of Mamba and Transformer within an advanced SE architecture (excluding CL and PCS).}
\label{table_unibi}
% \vspace{0.5em}
\vspace{0.8em}
\centering
% \begin{tabular*}{\linewidth}{l||lllll}
\begin{tabular*}{\linewidth}{@{\extracolsep{\fill}}l||lllll}
% \begin{tabular*}{8cm}{@{\extracolsep{\fill}} l || l l l l l}

\specialrule{1pt}{1pt}{1pt}
                  & PESQ  & STOI & FLOPs & Parameters \\ 
\hline
noisy-speech             & 1.97 & 0.92 &   -- &   -- \\
\hline

% Conform (Uni)           & XX & XXX  & No & XXX
% Mamba (Uni 2s)           & 3.17 & xxxx  & No & 1.41M   \\
% Mamba (Uni 3s)           & 3.29 & xxxx  & No & 1.41M   \\

Conformer           & 3.50 & 0.96  & 74.29G & 2.05M \\
Mamba (Uni)           & 3.29 & 0.95  & \textbf{53.09G} & \textbf{1.41M}   \\
% Mamba (Bi)           & \textbf{3.53} & \textbf{0.96}  & 65.46G & 2.25M   \\
Mamba (Bi)           & \textbf{3.52} & 0.96  & 65.46G & 2.26M   \\
\specialrule{1pt}{1pt}{1pt}
\end{tabular*}
% \vspace{-4mm}
\end{table}

%In this section, we present the dataset used to assess our approach and the experimental setup.

\subsection{Dataset}
\label{ssec:dataset}
We utilized the VoiceBank-DEMAND dataset~\cite{valentini2016investigating} for our study. This dataset comprises noisy speech recordings generated by mixing clean speech from the VoiceBank collection~\cite{veaux2013voice} with noise from the DEMAND dataset~\cite{thiemann2013diverse} and is widely used as one of the SE benchmark dataset. It encompasses 30 distinct speakers, with 28 speakers allocated for training and 2 for testing. Clean samples were mixed with noise samples at four signal-to-noise ratios (SNRs) during training ([0, 5, 10, 15] dB) and testing ([2.5, 7.5, 12.5, 17.5] dB). The training dataset comprised a total of 11,572 utterances, while the testing set included 824 utterances.

%This setup enables us to thoroughly examine the model's performance across diverse speech and noise combinations.
% \vspace{-2mm}
\subsection{Experimental Setup}
\label{ssec:setup}
Following \cite{Pascual2017}, all recordings were downsampled from 48 kHz to 16 kHz. The evaluation metrics used to assess the SE performance include: (i) Wide-band PESQ~\cite{Rix2001}, (ii) Prediction of the signal distortion (CSIG), (iii) Prediction of the background intrusiveness (CBAK), (iv) Prediction of the overall speech quality (COVL), and (v) STOI~\cite{Taal2011}. For all metrics, higher values  indicate better enhancement outcomes.

% In our experimental setup, we systematically evaluate various aspects to comprehensively assess the performance of the proposed SEMamba system\footnote{The detailed results and source codes of the model will be
% available for public access at \url{https://github.com/RoyChao19477/SEMamba}.}. First, we assess SEMamba-basic. Next, we assess Mamba with the more advanced model architecture. Finally, we examine the effectiveness of SEMamba-advanced with CL and PCS and evaluate its impact as pre-processing step in an ASR scenario as well.

In our experimental setup, we systematically evaluate various aspects to comprehensively assess the performance of the proposed SEMamba system\footnote{The detailed results and source codes of the model will be
available for public access at \url{https://github.com/RoyChao19477/SEMamba}.}. First, we assess SEMamba-basic. Next, we assess Mamba with the more advanced model architecture. Finally, we examine the effectiveness of SEMamba-advanced with CL and PCS and evaluate its impact as pre-processing step in an ASR scenario as well.

\begin{table}[]
\caption{The results of SEMamba and several well-known SE solutions on the VoiceBank-DEMAND dataset.}
\label{tab:se_comparison}
% \vspace{0.5em}
\vspace{0.8em}
\resizebox{\columnwidth}{!}{%
\begin{tabular}{@{}lccccc@{}}
\toprule
Model & PESQ & CSIG & CBAK & COVL & STOI \\ \midrule
noisy-speech             & 1.97 & 0.92 & 3.34 & 2.63 & 0.92 \\
\hline
SEGAN~\cite{Pascual2017} & 2.16 & 3.48 & 2.94 & 2.80 & -- \\
MetricGAN+ \cite{fu2021metricgan+}    & 3.15 & 4.14 & 3.16 & 3.64  & 0.93      \\ 
DPT \cite{dang2022dpt}  & 3.33 & 4.58 & 3.72 & 4.00 & 0.96      \\
CMGAN~\cite{CMGAN} & 3.41 & 4.63 & 3.94 & 4.12 & \textbf{0.96} \\
MP-SENet~\cite{mp_senet} & 3.50 & 4.73 & 3.95 & 4.22 & \textbf{0.96} \\ 
S4DSE~\cite{sun2024dual} & 2.55 & 3.94 & 3.00 & 3.23 & 0.93 \\
S4ND-UNet~\cite{ku2023multi} &  3.15  & 4.52  & 3.62  & 3.85 & -- \\
Spiking-S4~\cite{du2024spiking} &  3.39 & \textbf{4.92} &  2.64 & 4.31 & -- \\
\bottomrule
% Mamba(Uni-3s) & 3.29 & 4.XX & 3.XX & 4.XX & XXXX   \\

SEMamba (-CL) & 3.52 & 4.75 & \textbf{3.98} & 4.26 & \textbf{0.96} \\

% SEMamba (-CL+PCS) & 3.69 & 4.79 & 3.64 & 4.37 & \textbf{0.96} \\ \bottomrule

SEMamba & 3.55 & 4.77 & 3.95 & 4.29 & \textbf{0.96} \\
SEMamba (+PCS) & \textbf{3.69} & 4.79 & 3.63 & \textbf{4.37} & \textbf{0.96} \\ \bottomrule
\end{tabular}%
 }
% \vspace{-4mm}
\end{table}

% \vspace{-3mm}
\section{Experiments}
\label{sec:experiments}

% \vspace{-3mm}
\subsection{Experimental Results}
\label{sec:results}
Table \ref{table_uni} reports the experimental results with the basic SE architecture leveraging either the Mamba or Transformer block, as presented in Section II-A. Additionally, a non-causal configuration is also implemented as a support comparison. The results show that Mamba delivers comparable or superior performance to the Transformer in both causal and non-causal configurations while utilizing fewer FLOPs and parameters. Notably, the Mamba block reduces computational demands, needing 66.37\% fewer FLOPs and 60.22\% fewer parameters than the Transformer in causal configurations, and 53.09\% fewer FLOPs with 28.28\% fewer parameters in non-causal setups.

In the second set of experiments, we compare Mamba- and Transformer-based  advanced SE architectures. The results are listed in Table \ref{table_unibi}. The original MP-SENet uses a Conformer as the core component. In our experiments, Conformer and Transformer yield similar results. All three SE systems presented in Table \ref{table_unibi} are non-causal due to dilated DenseNet in feature encoder. Mamba (Bi) utilizes a bi-directional architecture, whereas Mamba (Uni) employs a uni-directional architecture. Comparing the results in Tables \ref{table_uni} and \ref{table_unibi}, we observe that integrating Mamba into an advanced model architecture notably improves performance. Additionally, Table \ref{table_unibi} reveals that Mamba not only achieves superior performance but also requires 11.88\% fewer FLOPs compared to Conformer.

% Both Conformer and Mamba (Bi) utilized a bi-directional architecture, while Mamba (Uni) employed a uni-directional architecture.
% \roychao{Moreover, SEMamba demonstrates a advantage over Conformer-based systems by managing longer input lengths under identical GPU VRAM constraints, as depicted in Fig \ref{fig:audio_length}.}

Third, we contrast the proposed SEMamba with several well-known SE methods on the VoiceBank-DEMAND dataset. The results are listed in Table \ref{tab:se_comparison}. For comparison, we include three SE methods based on the state-space model: S4DSE, S4ND-UNet, and Spiking-S4~\cite{sun2024dual, ku2023multi, du2024spiking}. 
%We also list the results without (denoted as -CL) and with consistency loss, as well as with (denoted as +PCS) and without PCS, for a detailed performance comparison.
The table also reports the results obtained with and without consistency loss (-CL), as well as with (+PCS) and without PCS, facilitating a detailed performance analysis. From Table \ref{tab:se_comparison}, we first note that SEMamba (-CL) yields performance comparable to MP-SENet. The major distinction between SEMamba (-CL) and MP-SENet is that SEMamba (-CL) utilizes Mamba as the core model, whereas MP-SENet employs Conformer. The results again confirm that Mamba yields similar evaluation scores. Notably, SEMamba reaches a high PESQ score of 3.55 on the VoiceBank-DEMAND dataset when consistency loss (CL) is included. Finally, by applying PCS, SEMamba’s attains a PESQ score of 3.69, achieving SOTA results on this dataset.

\begin{figure}[htb]
    \centering
    % \centerline{\includegraphics[width=14cm]{figures/simpleMamba.pdf}}
    % \centerline{\includegraphics[width=18.1cm]{figures/simpleMamba.pdf}}
    \centerline{\includegraphics[width=8.7cm]{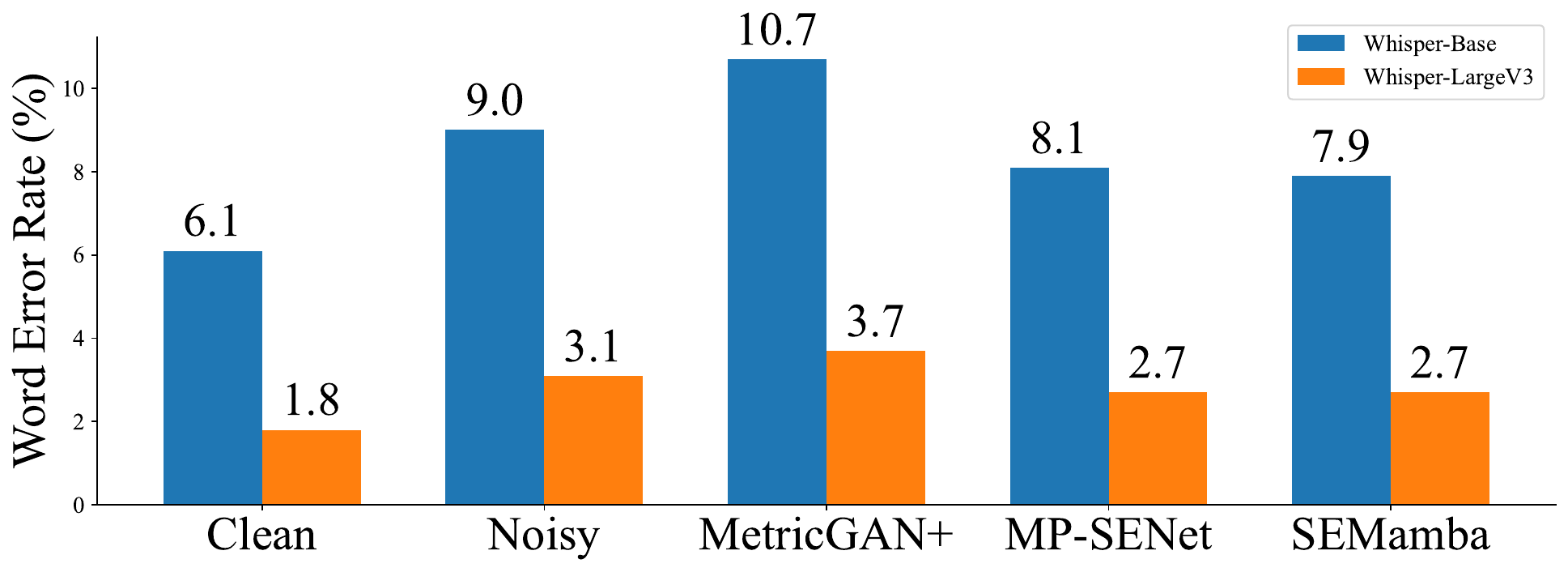}}
    \caption{Comparative analysis of WERs for SEMamba and related models on the VoiceBank-DEMAND dataset with Whisper ASR \cite{radford2023robust}.}
    \label{fig:audio_length}
\vspace{0mm}
\end{figure}

Finally, we have also tested ASR results using Whisper-Base and Whisper-LargeV3  on the test set of VoiceBank-DEMAND when the data are pre-processed by SEMamba-advanced\footnote{For the ASR performance evaluation, the SEMamba configuration we referred to is SEMamba-Advanced (PESQ=3.55). We conducted the ASR performance tests directly on this result without retraining. The testing script is at: \url{https://github.com/RoyChao19477/SEMamba}.}. For comparison, we tested related methods, MetricGAN+ and MP-SENet, as shown in Table \ref{tab:se_comparison} and both trained on VoiceBank-DEMAND. The results, presented in Fig. \ref{fig:audio_length}, show that SEMamba outperforms other methods in almost all conditions, achieving notable relative word error rate (WER) reductions of 12.22\% (from 9.0\% to 7.9\%) on Whisper-Base and 12.90\% (from 3.1\% to 2.7\%) on Whisper-LargeV3.

% Finally, we have also tested ASR results using Whisper-Base and Whisper-LargeV3  on the test set of VoiceBank-DEMAND when the data are pre-processed by SEMamba-advanced\footnote{The testing script is at: \url{https://github.com/RoyChao19477/SEMamba}.}. For comparison, we tested related methods, MetricGAN+ and MP-SENet, as shown in Table \ref{tab:se_comparison} and both trained on VoiceBank-DEMAND. The results, presented in Fig. \ref{fig:audio_length}, show that SEMamba outperforms other methods in almost all conditions, achieving notable relative word error rate (WER) reductions of 12.22\% (from 9.0\% to 7.9\%) on Whisper-Base and 12.90\% (from 3.1\% to 2.7\%) on Whisper-LargeV3.

% \vspace{-2mm}
\section{Conclusion}
\label{sec:conclusion}
This study explored Mamba, a novel SSM incorporating a selective mechanism, for tackling the SE task. Comprehensive investigations involving basic and advanced SE neural schemes and employing traditional signal-level and metric-oriented objective functions were carried out. Experimental results showed that Mamba holds noticeable promise in advancing SE performance. In fact, on the VoiceBank-DEMAND dataset, SEMamba with PCS achieved a SOTA PESQ score of 3.69. Furthermore, SEMamba used as a pre-processing step before ASR with advanced technologies allows, overall,  to achieve lower WERs. In future research, we aim to explore Mamba's potential in other speech generation tasks.

% \clearpage
% \vfill\pagebreak

\bibliographystyle{IEEEbib}
\bibliography{strings,refs}

\end{document}